# Spectroscopy of $Sr_2RuO_4$/Ru Junctions in Eutectic

H. Yaguchi [a], K. Takizawa [a], M. Kawamura [b,c], N. Kikugawa [a], Y. Maeno [a,d],
T. Meno [e], T. Akazaki [b], K. Semba [b] and H. Takayanagi [b]

[a] *Department of Physics, Graduate School of Science, Kyoto University, Kyoto 606-8502, Japan*
[b] *NTT Basic Research Laboratories, NTT Corporation, Atsugi, Kanagawa 243-0198*
[c] *RIKEN, Wako, Saitama 351-0198, Japan*
[d] *International Innovation Center, Kyoto University, Kyoto 606-8501, Japan*
[e] *NTT-ATN Corporation, Atsugi, Kanagawa 243-0018, Japan*

**Abstract.** We have investigated the tunnelling properties of the interface between superconducting $Sr_2RuO_4$ and a single Ru inclusion in eutectic. By using a micro-fabrication technique, we have made $Sr_2RuO_4$/Ru junctions on the eutectic system that consists of $Sr_2RuO_4$ and Ru micro-inclusions. Such a eutectic system exhibits surface superconductivity, called the 3-K phase. A zero bias conductance peak (ZBCP) was observed in the 3-K phase. We propose to use the onset of the ZBCP to delineate the phase boundary of a time-reversal symmetry breaking state.

It is now well established that the superconductor $Sr_2RuO_4$[1] is indeed a spin-triplet superconductor [2]. Importantly, an NMR experiment first revealed that the Knight shift of $^{17}O$ is not affected by the superconducting (SC) transition, indicative of the spin state of Cooper pairs is being triplet [3]. Subsequently, a muon spin relaxation measurement demonstrated that spontaneous magnetic moments accompany the SC transition, indicative of time reversal symmetry breaking in the SC phase [4]. Taken these results together, the basic form of the vector order parameter is constrained to be $\boldsymbol{d} = z(k_x + ik_y)$, corresponding to the chiral state. However, this basic form is too simplified to explain existing experimental results. The SC gap structure recently proposed based on detailed experiments on the angle dependence of the specific heat has successfully reconciled the discrepancy between the basic form of the order parameter and the existing experimental results [5].

Amongst the many interesting superconducting properties of $Sr_2RuO_4$, the enhancement of the SC transition temperature in $Sr_2RuO_4$-Ru eutectic is rather surprising [6]. This eutectic system, consisting of $Sr_2RuO_4$ and micron-size Ru inclusions, shows a broad SC transition with an onset of approximately 3 K, called the 3-K phase. Several experiments suggest that 3-K phase superconductivity is filamentary and occurs in the $Sr_2RuO_4$ side of the interface with Ru [7].

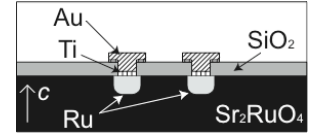

**FIGURE 1.** Schematic of the device fabricated on $Sr_2RuO_4$-Ru eutectic.

In order to measure the conductance at the interface between a single Ru micro inclusion and $Sr_2RuO_4$, we fabricated the device illustrated in Fig. 1 by using a micro-fabrication technique [8]. The device has a Ti/Au electrode directly attached to individual Ru micro-inclusions through 2 μm x 3 μm contact windows. The $SiO_2$ film is an insulating layer deposited on the polished ab-plane surface of $Sr_2RuO_4$-Ru eutectic. We were able to achieve good contacts only on Ru inclusions because of a non-superconducting layer forming on $Sr_2RuO_4$ surface. For actual measurements, therefore two $Sr_2RuO_4$/Ru junctions in series are inevitably involved.

Figure 2 shows the normalised differential conductance as a function of the bias voltage at temperatures between 0.3 K and 2.1 K. A clear zero bias conductance peak is seen, which is a hallmark of unconventional superconductivity and suggests a sign change of the order parameter on the Fermi surface [9]. As the ZBCP persists to a temperature well above

1.5 K, the ideal $T_c$ of bulk $Sr_2RuO_4$, the 3-K phase is responsible for the ZBCP. While the measurements involve *apparently* two junctions in series, we have confirmed only one of the two junctions *i.e.* a single interface of $Sr_2RuO_4$ and Ru is responsible for the ZBCP shown in Fig. 1 by the following way: We made three sets of measurements involving a third junction. Among them, one set of the measurements using the third junction gave rather featureless spectra, while the other two combinations gave the ZBCP. Thus we conclude that only one of the three junctions is responsible for the observed ZBCP. Whilst the spectra obtained in the present study are very similar to those previously reported on experiments using a break junction technique in Ref. 10, the width of the whole conductance peak in the present study is about the half that reported in Ref. 10. We suggest that this quantitative difference between the two works may be attributed effectively two junctions in series are involved in Ref. 10.

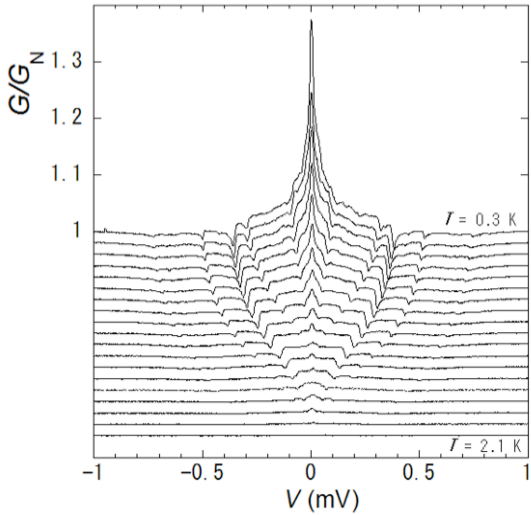

**FIGURE 2.** Normalised differential conductance at $H = 0$ as a function of the bias voltage at temperatures between 0.3 K and 2.1 K with a step of 0.1 K. A zero bias conductance peak is seen. Traces have been offset for clarity

We have also investigated the spectra in magnetic fields and observed the ZBCP. We have defined the onset magnetic field $H^*$ at which the ZBCP commences. Figure 3 is a plot of such $H^*$ as a function of temperature $T$, in addition to $H_{c2}(T)$ from Ref. 11. Noticeably, There is a clear difference between $H^*$ and $H_{c2}$ for $H$ // ab or $H \sim 0$ whilst those fields match with each other rather well for $H$ // c. Based on Sigrist and Monien's phenomenological theory, we suggest that the onset of the ZBCP $H^*$ corresponds to the onset of the chiral state. For $H$ // ab or $H \sim 0$, the nucleation of the state $d = zk_x$ occurs at the onset of the 3-K phase and the additional $k_y$ component with a relative phase of $\pi/2$, necessary for a time-reversal symmetry breaking state, is induced at a lower temperature. For $H$ // c, the onset of the 3-K phase already corresponds to a time-reversal symmetry breaking state due to the c-axis component of the applied field. A more detailed discussion with the help of Sigrist and Monien's theory [12] is described in Ref. 8.

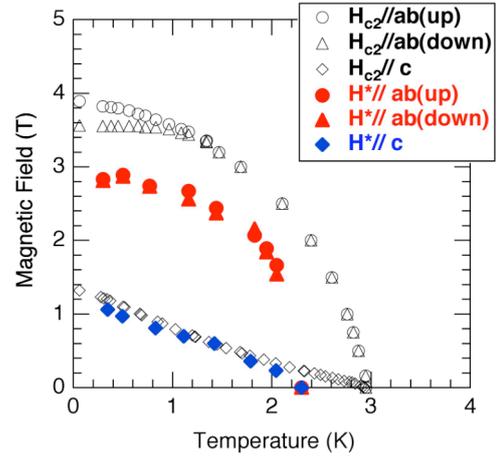

**FIGURE 3.** Onset of the ZBCP, $H^*$ and the upper critical field $H_{c2}$ as a function of temperature. There is a clear difference between $H^*$ and $H_{c2}$ for $H$ // ab or $H \sim 0$.

## ACKNOWLEDGEMENTS

This work was supported in part by a Grant-in-Aid for Scientific Research from the JSPS and 21COE program on "Center for Diversity and Universality in Physics" from the MEXT of Japan.